\documentclass[12pt,a4paper]{iopart}
\usepackage{graphicx}
\usepackage{amssymb}

\usepackage{braket}
\usepackage{bm}

\newcommand{\eqref}[1]{equation \eref{#1}}

\begin{document}

\title[Critical behavior of a chiral superfluid in a bipartite square lattice]{Critical behavior of a chiral superfluid in a bipartite square lattice}

\author{Junichi Okamoto$^{1,2}$, Wen-Min Huang$^{3,*}$, Robert H\"{o}ppner$^1$,  and Ludwig Mathey$^{1,2}$}

\address{$^1$ Zentrum f\"ur Optische Quantentechnologien and Institut f\"ur Laserphysik, Universit\"at Hamburg, 22761 Hamburg, Germany}
\address{$^2$ The Hamburg Centre for Ultrafast Imaging, Luruper Chaussee 149, 22761 Hamburg, Germany}
\address{$^3$ Department of Physics, National Chung-Hsing University, Taichung 40227, Taiwan}
\ead{$^*$wenmin@phys.nchu.edu.tw}

\begin{abstract}
We study the critical behavior of Bose-Einstein condensation in the second band of a bipartite optical square lattice in a renormalization group framework at one-loop order. Within our field theoretical representation of the system, we approximate the system as a two-component Bose gas in three dimensions. We demonstrate that the system is in a different universality class than the previously studied condensation in a frustrated triangular lattice due to an additional Umklapp scattering term, which stabilizes the chiral superfluid order at low temperatures. We derive the renormalization group flow of the system and show that this order persists in the low energy limit. Furthermore, the renormalization flow suggests that the phase transition from the thermal phase to the chiral superfluid state is first order. 
\end{abstract}

%
%
%

\section{Introduction}
Unconventional Bose-Einstein condensates (BECs) whose order parameter space is not simply the usual $U(1)$ symmetry have been extensively studied. Examples from the field of ultracold atoms include Floquet engineered Bose gases \cite{Struck2011, Struck2013, Parker2013, Clark2016} or spinor Bose gases \cite{Kawaguchi2012}, where the order parameter space has an additional Ising component or even more complex symmetry groups. The experimental realization of such systems in ultracold atomic systems are ideal to investigate phase transitions of those complex orders due to well-defined and tunable nature of these systems. 

Recently, BECs that break time-reversal (TR) symmetry have attracted increased attention from theorists \cite{Isacsson2005, Liu2006, Kuklov2006, Wu2006, Lim2008, Stojanovic2008, Wu2009, Lewenstein2011, Cai2011, Li2012,  Martikainen2012, Cai2012, Liu2013, Li2016} and from experimentalists \cite{Olschlager2011,Wirth2011, Soltan-Panahi2012, Olschlager2013, Kock2015}. According to Feynman's ``no-node" theorem \cite{feynman1998statistical}, such states cannot be a ground state of a conventional bosonic Hamiltonian with short-range interactions, since breaking TR symmetry inevitably leads to a wave function with a node in real space. Therefore, to create BECs without TR symmetry, the assumptions of the no-node theorem have to be circumvented. One approach uses a long-lived metastable state of ultracold bosons in bands of higher orbitals \cite{Wu2009}. Experimentally, a BEC in a $p$-band has been realized by first populating particles in a staggered pattern in a checkerboard lattice and then suddenly changing the potential shape. Due to the large anharmonicity in the energy spectrum, the life time of the metastable BEC is longer than 100ms \cite{Stojanovic2008, Wirth2011}. Since this BEC is not a ground state, breaking TR symmetry is not in contradiction with the no-node theorem. Indeed, a complex coherent superposition of the two $p$-band condensates (i.e., $p_x \pm i p_y$ order) that breaks TR symmetry has been realized by carefully tuning the lattice parameters \cite{Wirth2011, Olschlager2013, Kock2015}. Since such a state hosts spatially staggered orbital currents, it is dubbed a chiral superfluid. Similar $p_x \pm i p_y$ paring has been proposed for the A phase of superfluid $^3$H \cite{Leggett1975, dobbs2000helium} and for Sr$_2$RuO$_4$ \cite{Mackenzie2003}.

In this paper, we investigate the stability of the chiral condensate and its critical behaviors by a renormalization group (RG) analysis. The analysis addresses the competition of chiral and non-chiral condensation, and the critical behavior. While the chiral superfluid has been confirmed experimentally, it is still important to know how stable and general the state is. In particular, since the energies of the competing non-chiral BEC and of the chiral BEC are close at the mean-field level, it is not trivial which of the two BECs becomes dominant at low temperatures. We find that the stable condition of the chiral BEC is always preserved at low energy scales, and the transition is expected to be first-order. 

The paper is organized as follows. In \sref{EFT} we develop the field theoretical description of the mixed orbital model in a bipartite optical square lattice in the low energy limit. \Sref{MF} is devoted to a mean-field analysis of the effective model. In \sref{rg}, we study the critical behavior of the model in the framework of a one-loop RG calculation. In \sref{conc}, we conclude. The details of calculations not covered in the main texts are summarized in appendix.


\section{Effective field theory}\label{EFT}
  
\begin{figure*}[t!]\center
\includegraphics[height=7cm]{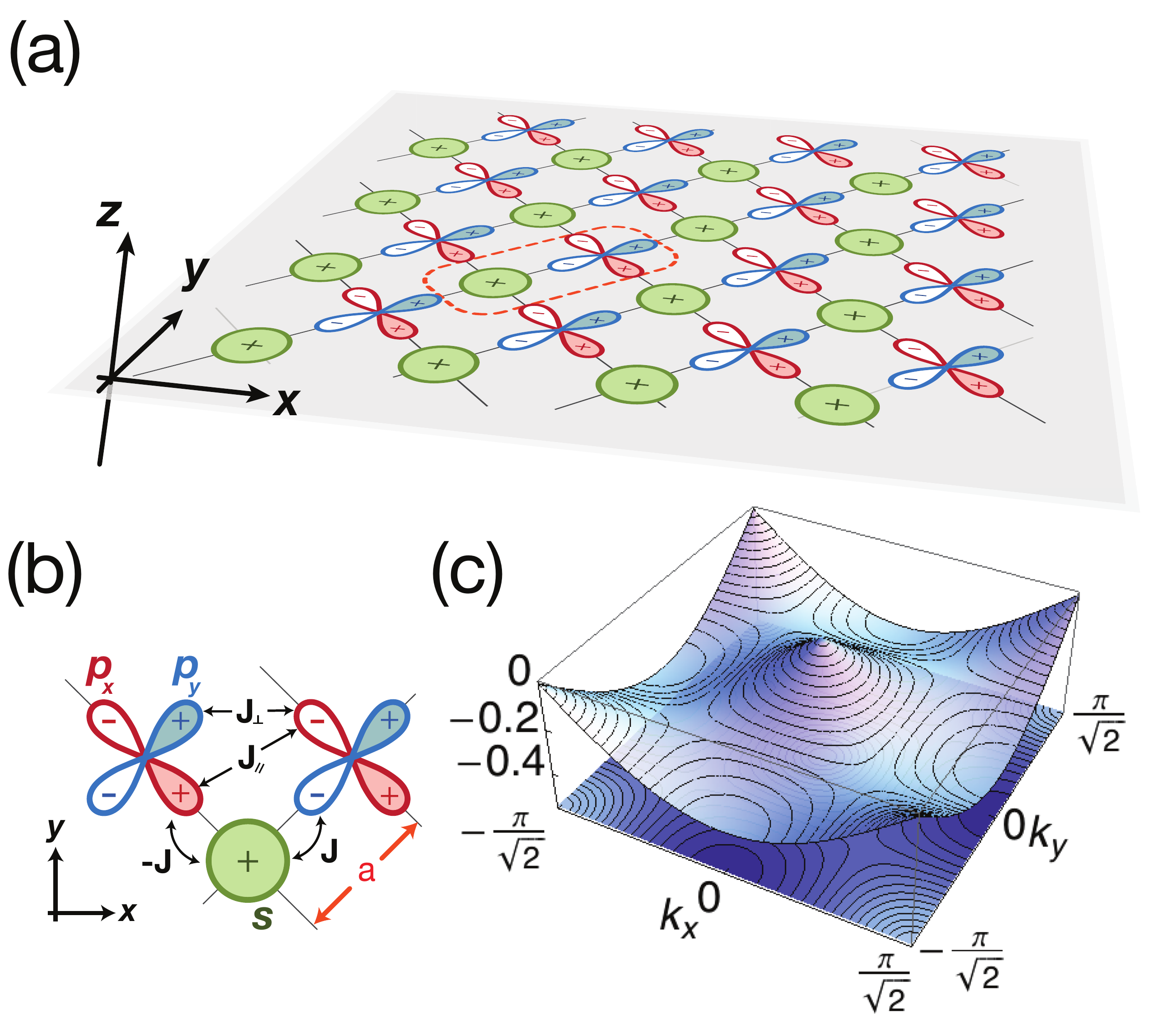}
\caption{(a) A bipartite optical square lattice is sketched in the $xy$ plane with $s$-and $p$-orbitals arranged in a chequerboard pattern. Along the $z$-direction, bosons move freely. (b) The hopping amplitudes between orbitals on different sites, and the lattice constant $a$ are illustrated. (c) The lowest band of the bipartite optical square lattice in units of $E_{\rm rec}$ is plotted for the hopping amplitudes $J=0.13E_{\rm rec}$, $J_{\perp}=0$ and $J_{\|}=0$; these choices are motivated by the experimental setup in \cite{Olschlager2013}. }
\label{lattice}
\end{figure*}

The system that we consider here is sketched in \fref{lattice}(a) and described by the Hamiltonian, 
\begin{eqnarray}\label{h0}
H_{0}= \int dz
\Bigg\{ \sum_{\bm{r},i} b^{\dagger}_i(\bm{r},z)\left(\frac{-\hbar^{2}\partial^2_z}{2m_0}-\mu_{\rm 3D}\right) b_i(\bm{r},z)+H^{\rm xy}_0\Bigg\},
\end{eqnarray}
with the tight-binding model of a bipartite optical square lattice
\begin{eqnarray}\label{h0xy}
\nonumber H^{\rm xy}_0&=&J\sum_{\bm{r}} \Big[b_{1}^{\dag}(\bm{r},z)b_{2}(\bm{r}+\bm{d}_1,z)+b_{1}^{\dag}(\bm{r},z)b_{3}(\bm{r}+\bm{d}_2,z)\\&&
-b_{1}^{\dag}(\bm{r},z)b_{2}(\bm{r}-\bm{d}_1,z)-b_{1}^{\dag}(\bm{r},z)b_{3}(\bm{r}-\bm{d}_2,z)+{\rm h.c.}\Big].
\end{eqnarray}
Here $\bm{r} = \sqrt{2} a (n_x, n_y)$ with $n_{x,y} \in \mathbb{Z}$, $\bm{d}_{1/2}=(a/\sqrt{2},\pm a/\sqrt{2})$. $b_i(\bm{r},z)$ with $i=1,2,3$ represent the annihilation operators of bosons at the $s$, $p_x$ and $p_y$ orbitals respectively. We assume that bosons move freely along the $z$-direction. The hopping amplitudes between neighboring $p$-orbitals, $J_{\parallel}$ and $J_{\perp}$, are set to be zero for simplicity (see appendix for a more general discussion). Converting the orbital representation into a band representation, we obtain three bands. The metastable BEC in experiments is loaded in the lowest band, whose dispersion is given by
\begin{eqnarray}
\epsilon(\bm{k},k_z)=-2J\sqrt{1-\cos\left(\sqrt{2}k_x\right)\cos\left(\sqrt{2}k_y\right)}+\frac{\hbar^2 k_z^2}{2m_0},
\end{eqnarray}
where $\bm{k}=(k_x,k_y)$ and we set $a=1$ in the following calculations. We illustrate the lowest band in momentum space in \fref{lattice}(c). We note that there are two energetic minima at $\bm{k}_1=(\pi/\sqrt{2},0)$ and $\bm{k}_2= (0,\pi/\sqrt{2})$. These energetic minima are degenerate, thus giving rise to the $\mathbb{Z}_2$ symmetry of the noninteracting Hamiltonian. 

At low temperatures, bosons predominantly occupy momentum states near the two minima, and then condense below a critical temperature. To describe the critical behavior, we expand the bosonic operators near the two minima as
\begin{eqnarray}\label{field}
\nonumber b_{\alpha}(\bm{r},z)&=&\frac{1}{\sqrt{N}}\sum_{\bm{k}}e^{i\bm{k}\cdot\bm{r}}u_{\alpha}({\bm{k}})\phi(\bm{k},z)\\
\nonumber&\simeq&\frac{2a^2}{\sqrt{N}}\hspace{-0.1cm}\sum_{j=1,2}e^{i\bm{k}_j\cdot\bm{r}}u_{\alpha j}\hspace{-0.1cm}\int_{|\bm{q}_j|<\Lambda_q}\hspace{-0.2cm}\frac{d^2\bm{q}_j}{4\pi^2}~e^{i\bm{q}_j\cdot\bm{r}}\phi_j(\bm{q}_j,z)\\
&\equiv&\sqrt{2}a\sum_{j=1,2}\psi_j(\bm{r},z)u_{\alpha j}~e^{i\bm{k}_j\cdot\bm{r}},
\end{eqnarray}
where $\psi_j(\bm{r},z)\equiv\frac{\sqrt{2}a}{\sqrt{N}}\int_{|\bm{q}_j|<\Lambda_q}\frac{d^2\bm{q}_j}{4\pi^2}e^{i\bm{q}_j\cdot\bm{r}}\phi_j(\bm{q}_j,z)$ with $\phi(\bm{k}_j+\bm{q}_j,z)\equiv\phi_j(\bm{q}_j,z)$. $\Lambda_q$ is the momentum cut-off, and $N$ is the number of the unit cells. The kernel $u_{\alpha}(\bm{k}_j)=u_{\alpha j}$ represents the projection of the wave function of orbital $\alpha$ on the wave function of the lowest band in the vicinity of the minimum $\bm{k}_j$. These are given by $(u_{11},u_{12})=(-i/\sqrt{2},i/\sqrt{2})$, $(u_{21},u_{22})=(1/2, -1/2)$ and $(u_{31},u_{32})=(1/2,1/2)$. We use the field decomposition to approximate the full Hamiltonian of \eqref{h0} by an effective Hamiltonian with two components, 
\begin{eqnarray}
H_0^{\rm eff}=\sum_{j=1,2}\int d^3\bm{R}\ \ \psi^{\dag}_j(\bm{R})\left[\frac{-\hbar^2}{2m^*}\nabla_{\bm R}^2-\mu_j\right]\psi_j(\bm{R}),
\end{eqnarray}
where $\bm{R}=(\bm{r},z)$, $\mu_{j}$ being the chemical potential of the $j$-th component and the effective mass being $m^*=(m_{\rm xy}^2m_0)^{1/3}$. As an example, we describe the experimental parameters of  \cite{Wirth2011}; for $^{87}$Rb atoms and for typical laser intensities that were used, we have $m_0\simeq0.2m_{\rm xy}$ and $m_{\rm xy}=2\sqrt{2}\hbar^2/\left(\lambda_L^2J\right)$ with the laser wavelength $\lambda_L=1064{\rm nm}$ and $J/E_{\rm rec}=0.13$. We note that to simplify the RG analysis we use an isotropic effective model, and the momentum cut-off in the field decomposition sets the energy cut-off of the effective Hamiltonian as $\epsilon_{\Lambda}=\hbar^2\Lambda_{q}^2/2m^*$.

We further consider the on-site interaction; see \cite{Olschlager2013}, which gives the following terms,
\begin{eqnarray}\label{oHI}
\nonumber H_I&=& \int dz\sum_{\bm{r}}\Bigg\{\frac{U_s}{2}n_s(\bm{R})\left[n_s(\bm{R})-1\right]\\
&&+\frac{U_p}{2}n_p(\bm{R})\left[n_p(\bm{R})-1 \right]-\frac{U'_p}{2} \left[ L_z^2(\bm{R}) - n_p(\bm{R}) \right] \Bigg\},
\end{eqnarray}
where $n_s =b_1^{\dag} b_1$, $n_p=b_2^{\dag}b_2+b_3^{\dag}b_3$, and $L_z=i\left(b^{\dag}_2b_3-b^{\dag}_3b_2\right)$ being an angular momentum operator. $U_s$ is the on-site interaction among $s$-orbitals. $U_p$ and $U'_p$ are intra- and inter-orbital interactions among $p$-orbitals. In the tight-binding approximation, the strength of the on-site interactions can be calculated from the contact interaction and the Wannier functions~\cite{Li2016}; for the details, see appendix. In the standard harmonic approximation, the on-site interactions follow $U'_p= U_p/3$. The precise ratio between $U_s$ and $U_p$ depends on the depth of the optical potential since the harmonic frequency of the $s$-orbital sites are different from the one of the $p$-orbital sites. For a moderately deep potential, we find $U_s \sim U_p$. 

Within the field-theory approximation \eref{field}, we represent the effective interaction as 
\begin{eqnarray}\label{Hieff}
&&\nonumber\hspace{-1.8cm} H^{\rm eff}_I=\int d^3\bm{R}\Bigg\{\frac{\tilde{g}_1}{2}\psi_1^{\dag}(\bm{R})\psi_1^{\dag}(\bm{R})\psi_1(\bm{R})\psi_1(\bm{R})+\frac{\tilde{g}_2}{2}\psi_2^{\dag}(\bm{R})\psi_2^{\dag}(\bm{R})\psi_2(\bm{R})\psi_2(\bm{R})\\
&&\hspace{-0.8cm} +\tilde{g}_{12}\psi_1^{\dag}(\bm{R})\psi_{2}^{\dag}(\bm{R})\psi_{2}(\bm{R})\psi_{1}(\bm{R})+\frac{\tilde{g}_u}{2}\Big[\psi_1^{\dag}(\bm{R})\psi_1^{\dag}(\bm{R})\psi_{2}(\bm{R})\psi_{2}(\bm{R})+{\rm H.c.}\Big]\Bigg\},
\end{eqnarray}
where $\tilde{g}_{j}$ ($j=1,2$) is the intra-component interaction and  $\tilde{g}_{12}$ is the inter-component one. In this expression, there is an additional term with the interaction strength $\tilde{g}_u$, which is an Umklapp term. This additional scattering process is not present in the previously studied triangular lattice system \cite{Antonenko1994, Kawamura1998, Janzen2016}, which demonstrates that these two systems are in different universality classes. Our model is more general in the sense that three coupling constants flow independently under the renormalization equations. The Umklapp interaction allows interchange of bosons between the two components by lattice assisted collisions. In other words, the effective interaction only enforces conservation of the total boson number, in stead of the boson number of each component as, for instance, in the frustrated triangular optical lattice~\cite{Janzen2016}. Equivalently, we only have one global $U(1)$ symmetry, instead of two $U(1)$ symmetries for each component. The bare values of the coupling constants in terms of $U_s$, $U_p$ and $U'_p$ are:
\begin{eqnarray}\label{effective_int}
\tilde{g}_{1/2} =  \frac{2 U_s+U_p + 3 U_p'}{8}, \ \tilde{g}_{12}= \frac{2U_s+U_p-U_p'}{4}, \ \tilde{g}_{u} = \frac{1}{2}\tilde{g}_{12}.
\end{eqnarray}

The full symmetry of the effective action is $U(1)\times \mathbb{Z}_2 \times \Theta$, where the $U(1)$ symmetry corresponds to the invariance of the model under the global phase shift for both components, $\mathbb{Z}_2$ is the exchange of the two components, and $\Theta$ is the time-reversal symmetry.


\section{Mean-field theory}\label{MF}
Before proceeding to the RG analysis, we study the ground state for the bare interactions within a zero-temperature approach. We assume that the bosons perfectly condense at the two energetic minima so that a many-body trial wave function is represented as,
\begin{eqnarray}\label{trial}
\left|\Psi\right\rangle_{\theta,\phi}=\frac{1}{\sqrt{N!}}\left[\cos\theta \psi_{1}^{\dag}+e^{i\phi}\sin\theta \psi_{2}^{\dag}\right]^{N}\left|0,0\right\rangle,
\end{eqnarray}
where $N$ is the total number of bosons, and $\left|m,n\right\rangle$ stands for bosons' occupation numbers $m(n)$ at momentum $\bm{k}_1(\bm{k}_2)$  respectively. We will use the angles $\theta$ and $\phi$ as variational parameters. $\theta$ determines the relative population of the two minima and $\phi$ denotes the relative phase of the two-component condensates. Using the trial wave function, we compute the energy of the interacting effective Hamiltonian \eref{Hieff} as
\begin{eqnarray}\label{energy}
&&\nonumber\hspace{-1cm}\left\langle H^{\rm eff}_I\right\rangle_{\theta,\phi}=N(N-1)\Bigg\{\frac{\tilde{g}_1}{2}\cos^4\theta+\frac{\tilde{g}_2}{2}\sin^4\theta+\tilde{g}_{12}\cos^2\theta\sin^2\theta\\
&&\hspace{2cm}+\tilde{g}_u\cos^2\theta\sin^2\theta\cos2\phi\Bigg\}.
\end{eqnarray}
First we note that for the frustrated triangular optical lattice, the Umklapp interaction does not occur ($\tilde{g}_u = 0$), and we have $\tilde{g}_1=\tilde{g}_2 \approx 2\tilde{g}_{12}$, which follows from the common origin of these terms, i.e., the contact interaction between the atoms. In this case, the minimum of \eqref{energy} occurs at $\theta=0$ or $\pi/2$. From \eqref{trial}, this means that bosons will condense in one of the energetic minima to break the $\mathbb{Z}_2$ symmetry \cite{Janzen2016}. However, in the square bipartite lattice that we study in this paper, the Umklapp interaction $\tilde{g}_u>0$ exists due to the bare on-site repulsive interactions $U_s \sim U_p \gg U'_p >0$. In this case, another energetic minimum may appear at $(\phi, \theta)=(\pm\pi/2, \pi/4)$ in \eqref{energy}; this corresponds to a chiral superfluid state $\left|\Psi\right\rangle= \left[\psi_{1}^{\dag}\pm i \psi_{2}^{\dag}\right]^{N}\left|0,0\right\rangle /\sqrt{2N!}$ given by a complex coherent superposition of two single particle states. This state breaks the time-reversal symmetry $\Theta$, i.e., the chiral $\mathbb{Z}_2$ symmetry, in addition to the $U(1)$ continuous symmetry of the phase (The situation is similar to the fully frustrated XY models \cite{Villain1977}). Comparing this to the single condensate at $\theta = 0$, we find that the chiral superfluid state occurs when
\begin{eqnarray}\label{chiralC}
G_1 \equiv \tilde{g}_0-\tilde{g}_{12}+\tilde{g}_u>0,
\end{eqnarray}
where we set $\tilde{g}_1=\tilde{g}_2=\tilde{g}_0$. The above condition is also discussed in \cite{Li2016}. In terms of the interaction parameters in \eqref{oHI}, we find $\tilde{g}_0-\tilde{g}_{12}+\tilde{g}_u = U'_p/2 >0$, and thus a chiral superfluid order occurs for any repulsive interaction. Even if we include non-zero values of $J_{\perp}$ and $J_{\parallel}$, we find that the condition is still satisfied as long as the energetic minima are located at $\bm{k}_1=(\pi/\sqrt{2},0)$ and $\bm{k}_2= (0,\pi/\sqrt{2})$ (see appendix). However, the energy difference between the normal and chiral condensates is of the order of $\sim U_p'$ at the mean-field level, and at low temperatures the coupling constants get renormalized under the RG flow. Then it is nontrivial which superfluid order emerges at low temperatures. To study this problem more systematically, and to study the critical behavior in the low-energy limit, we apply the renormalization group method in the next section.   

\section{One-loop renormalization group method}\label{rg}
Following the RG method employed in~\cite{Janzen2016} and ignoring quantum fluctuations, we calculate the RG equations at one-loop order as \cite{stoof2008ultracold}, 
\numparts
\begin{eqnarray}
\frac{d\mu_{\Lambda}}{dl}&=&2\mu_{\Lambda}-T_{\Lambda} F(\mu_{\Lambda})\left(2 g_0 + g_{12}\right),\label{muFlow}\\
\frac{dg_0}{dl}&=& \epsilon g_0-T_{\Lambda}F(\mu_{\Lambda})^2 \left( 5g_0^2 + g_{12}^2 + g_u^2 \right), \label{g0Flow}\\
\frac{dg_{12}}{dl}&=& \epsilon g_{12}-T_{\Lambda}F(\mu_{\Lambda})^2 \left( 4g_0g_{12}+2g_{12}^2+4g_u^2\right), \label{g12Flow}\\
\frac{dg_{u}}{dl}&=& \epsilon g_u-T_{\Lambda}F(\mu_{\Lambda})^2 \left(2g_0g_{u}+4g_{12}g_{u}\right),\label{guFlow}
\end{eqnarray}
\endnumparts
where $l= \ln(\Lambda_q/\Lambda_b)$ is the logarithm of the ratio between the bare momentum cutoff $\Lambda_q$ and the running cutoff $\Lambda_b$. Here $\epsilon = 4-d$ with $d$ being the spatial dimension of the system, and $\epsilon = 1$ for our three-dimensional model. We have also defined dimensionless parameters,  $\mu_{\Lambda}=\mu_1/\epsilon_{\Lambda}=\mu_2/\epsilon_{\Lambda}$, $T_{\Lambda}=k_BT/\epsilon_{\Lambda}$, $g_i=\tilde{g}_i \Lambda_q^3 /(2\pi^2\epsilon_{\Lambda})$, and $F(\mu_{\Lambda}) = 1/(1-\mu_{\Lambda})$. If $g_u=0$, the flow equations become the ones of the two-component $\phi^4$-theory~\cite{Janzen2016}. In the critical regime, $\mu_{\Lambda} \ll 1$, where we can approximate $F(\mu_{\Lambda}) \approx 1$, \footnote{This corresponds to the lowest order of the $\epsilon$-expansion \cite{Janzen2016, cardy1996scaling, Wilson1972, Domany1977}. The effect of higher order contributions to the fixed points is discussed in appendix.} the RG equations exhibit four fixed points, except the trivial one, $(\mu^*_\Lambda,g^*_0,g^*_{12},g^*_u)=(0,0,0,0)$,
\numparts
\begin{eqnarray}
\left(\mu^*_\Lambda,g^*_0,g^*_{12},g^*_u\right)&=&\left(\frac{1}{5},\frac{1}{5T_{\Lambda}},0,0\right),\label{fp1}\\
\left(\mu^*_\Lambda,g^*_0,g^*_{12},g^*_u\right)&=&\left(\frac{1}{4},\frac{1}{6T_{\Lambda}},\frac{1}{6T_{\Lambda}},0\right),\label{fp2}\\
\left(\mu^*_\Lambda,g^*_0,g^*_{12},g^*_u\right)&=&\left(\frac{1}{5},\frac{1}{10T_{\Lambda}},\frac{1}{5T_{\Lambda}},\frac{1}{10T_{\Lambda}}\right),\label{fp3}\\
\left(\mu^*_\Lambda,g^*_0,g^*_{12},g^*_u\right)&=&\left(\frac{1}{5},\frac{1}{10T_{\Lambda}},\frac{1}{5T_{\Lambda}},-\frac{1}{10T_{\Lambda}}\right).\label{fp4}
\end{eqnarray}
\label{fp}
\endnumparts
All these fixed points are unstable, indicating that the system undergoes a first-order transition. In \fref{flowD}(a), we show the fixed points as red points for $g_u\geq0$. 

\subsection{Basic structures of RG equations}
While the full RG equations are complicated and can only be solved numerically,  the basic structure of the equations give useful insight in the RG flow. In particular, we find three separatrix surfaces of the RG equations. RG flow cannot pass through these surfaces, and therefore the asymptotic behavior of the RG flow is severely constrained by the initial condition. 

\begin{figure}[t!]\center
\includegraphics[height=9cm]{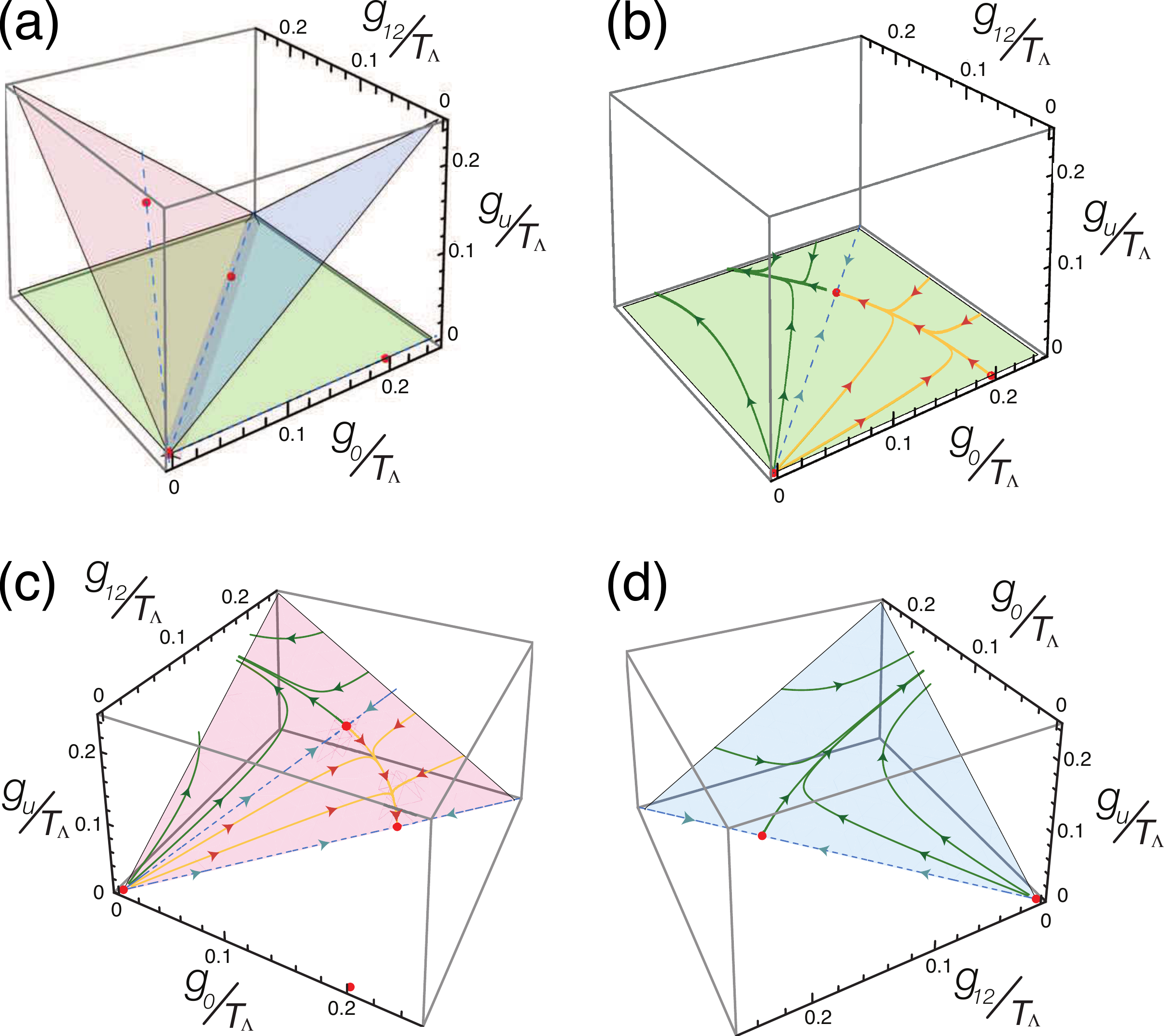}
\caption{(a) Three planes separating the RG flow in the $g_0$-$g_{12}$-$g_u$ space for $g_u>0$. The four fixed points are plotted as red dots. (b) The RG flow on the $g_u=0$ plane, which are also introduced in \cite{Janzen2016}. (c) RG flows on the $G_1 = 0$ plane. (d) RG flows on the $G_2 = 0$ plane.}
\label{flowD}
\end{figure}

The first separatrix is $g_u=0$; the formal solution of \eqref{guFlow} is
\begin{eqnarray}\label{guSol}
g_u(l) = g_u (0)  \exp \left\{ \int_0^l dl' \left[ 1-T_{\Lambda}F(\mu_{\Lambda})^2 \left(2 g_0 + 4  g_{12}\right) \right] \right\},
\end{eqnarray}
which explicitly shows that $g_u$ does not change signs under the RG. In \fref{flowD}(b), we show the flow diagram on the $g_u=0$ surface, which is also introduced in \cite{Janzen2016}. The two fixed points $(g^*_0,g^*_{12},g^*_u)=(0,0,0)$ and $(1/(5T_{\Lambda}),0,0)$ are unstable, and the one at $(1/(6T_{\Lambda}),1/(6T_{\Lambda}),0)$ is marginally unstable.  

The second separatrix surface is $G_1\equiv g_0-g_{12}+g_u = 0$. The RG equation for  $G_1$ is 
\begin{eqnarray}\label{G1}
\frac{dG_1}{dl}=G_1-T_{\Lambda}F(\mu_{\Lambda})^2G_1(5g_0+g_{12}-3g_u).
\end{eqnarray}
We emphasize that $G_1 > 0$ coincides with the mean-field stable condition for a chiral superfluid, \eref{chiralC}, and therefore, for general repulsive interactions that give $G_1 >0$ as an initial condition, a chiral superfluid order always persists in the low-energy limit. In \fref{flowD}(c), we illustrate the flow diagrams on the $G_1=0$ plane for $g_u>0$. There are two unstable fixed points, \eref{fp2} and \eref{fp3}, on the plane, in addition to the trivial one $(g^*_0,g^*_{12},g^*_u)=(0,0,0)$. In particular the ray from $(g^*_0,g^*_{12},g^*_u)=(0,0,0)$ to $\left(1/10T_{\Lambda},1/5T_{\Lambda},1/10T_{\Lambda}\right)$ separates the RG flows into two parts. For initial conditions $g_0>g_u$ on the $G_1=0$ plane, we find that system eventually flows into the fixed point $(g^*_0,g^*_{12},g^*_u)=\left(1/6T_{\Lambda},1/6T_{\Lambda},0\right)$, where the system is reduced to the standard two-component $\phi^4$ theory. However, if the initial conditions deviate from the $G_1=0$ plane by an arbitrarily small amount, $g_u$ will grow to a large positive value, and thus the fixed point \eref{fp2} is actually unstable. For initial conditions $g_0 < g_u$ on the $G_1=0$ plane, the flow first approaches $\left(1/10T_{\Lambda},1/5T_{\Lambda},1/10T_{\Lambda}\right)$ and then runs away to larger positive values of $g_u$.


Finally the third separatrix surface is $G_2 \equiv g_0-g_{12}-g_u =0$, and its RG equation is
\begin{eqnarray}
\frac{dG_2}{dl}=G_2-T_{\Lambda}F(\mu_{\Lambda})^2G_2(5g_0+g_{12}+3g_u).
\end{eqnarray}
The flow equation is similar to \eqref{G1}, and it guarantees that no flow passes through the $G_2=0$ plane. As illustrated in \fref{flowD}(d), we find that the Umklapp interaction $g_u$ always grows up on the $G_2=0$ plane. The increase of $g_u$ is also found when the initial couplings deviate from the $G_2$ plane as we discuss below.

\subsection{RG flows for the effective model}
Now let us analyze the RG equations for the relevant parameter regime of our model. For our effective model, we have $U_s \sim U_p \sim U'_p/3 >0$, and the RG flow is constrained to the space of $g_u>0$, $G_1 >0$ and $G_2 <0$. Due to this constraint, as we will show below, the possible phases of our effective model are either the thermal gas phase ($\mu_{\Lambda} \rightarrow - \infty$) or the chiral superfluid phase ($\mu_{\Lambda} \rightarrow + 1$). The phases are determined by the non-universal nature of the RG flows and initial conditions. 

\begin{figure}[!bt]\center
\includegraphics[height=5cm]{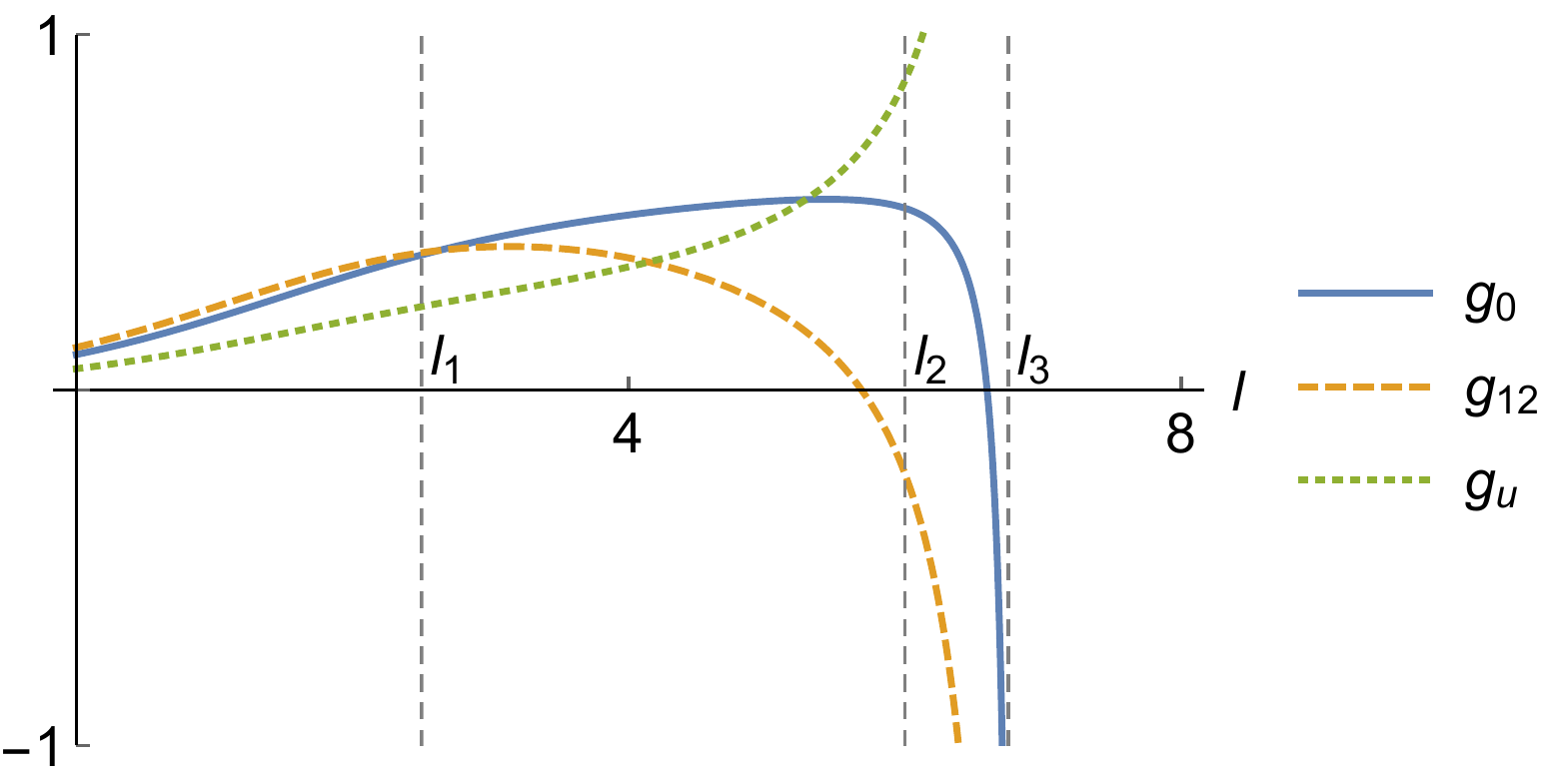}
\caption{A schematic typical RG flow for the three coupling constants when $\mu_{\Lambda}$ remains positive. The characteristic length scales that separate the behavior of the flow are also plotted. }
\label{RG_flow}
\end{figure}

First, when $\mu_{\Lambda}$ remains positive, a typical flow of coupling constants behaves as in \fref{RG_flow}. For small positive initial interactions, a typical flow has monotonically increasing $g_u$, which can also be seen from \eqref{guSol}. For the evolution of $g_0$ and $g_{12}$, there are three regimes that we can characterize:
\begin{enumerate}
\item $0 < l < l_1$ (the initial regime): the linear terms in RG equations are dominant, and $g_0$ and $g_{12}$ gradually increase.
\item $l_1 < l < l_2$ (the intermediate regime): the quadratic terms become more important with increasing $g_u$, which eventually make $g_0$ and $g_{12}$ negative. 
\item $l_2 < l< l_3$ (the asymptotic regime): the quadratic terms give asymptotically diverging behaviors, while the coupling constants are still smaller than the unity. 
\end{enumerate}
Of course, we should stop the RG flow before any of the coupling constant becomes order of unity near $l_3$, above which the cubic terms become dominant. In the asymptotic regime, $\mu_{\Lambda}$ approaches one, and the quadratic terms in the RG equations become dominant. Ignoring the linear terms, the flow can be analyzed by an ansatz $g_i (l) = k \bar{g}_i/(1- kl)$, where $k$ is the inverse length scale at which these coupling constants diverge \cite{Balents1996}. We find that the only asymptotic flow constrained in the space of $g_u>0$, $G_1 >0$ and $G_2 <0$ is given by 
\begin{equation}
(\bar{g}_0, \bar{g}_{12}, \bar{g}_{u}) = \frac{1}{T_{\Lambda}F(\mu_{\Lambda})^2}\left( -\frac{1}{10}, -\frac{1}{5}, \frac{1}{10} \right).
\label{asymptotic} 
\end{equation}
This implies that $G_1 $ steadily increases and thus stabilizes the chiral superfluid order, \eref{chiralC}. At the same time, the quartic interactions $g_0$ and $g_{12}$ are renormalized to negative values. This indicates the breakdown of the quartic effective field theory, and we need to include higher order interactions such as
\begin{equation}
g_6 \int d^3\bm{R} \left\{ |\psi_1 (\bm{R})|^6 + |\psi_2 (\bm{R})|^6 \right\},
\end{equation}
which is generated from three quartic vertices after one-loop renormalization. We note that at tree-level the $g_6$ contribution is marginal, and stabilizes the system \cite{Janzen2016}. At one-loop order, one contribution to the RG equation for $g_6$ is $d g_6/ dl \simeq  - 24 T_{\Lambda} F(\mu_{\Lambda})^2 g_{0} g_6$. Since $g_0$ flows to negative values, the above contribution further stabilizes the system. 

Second, when the chemical potential becomes negative, $F(\mu_{\Lambda})=1/(1-\mu_{\Lambda})$ gets suppressed, making the linear terms in the RG equations more dominant. Therefore, typical RG flows give a simple scaling behavior, $(\mu_{\Lambda},g_i ) \sim \left(- e^{2l},  e^l \right)$. This corresponds to the thermal gas phase without condensation.

\subsection{Phase diagram}
\begin{figure}[t!]\center
\includegraphics[height=6cm]{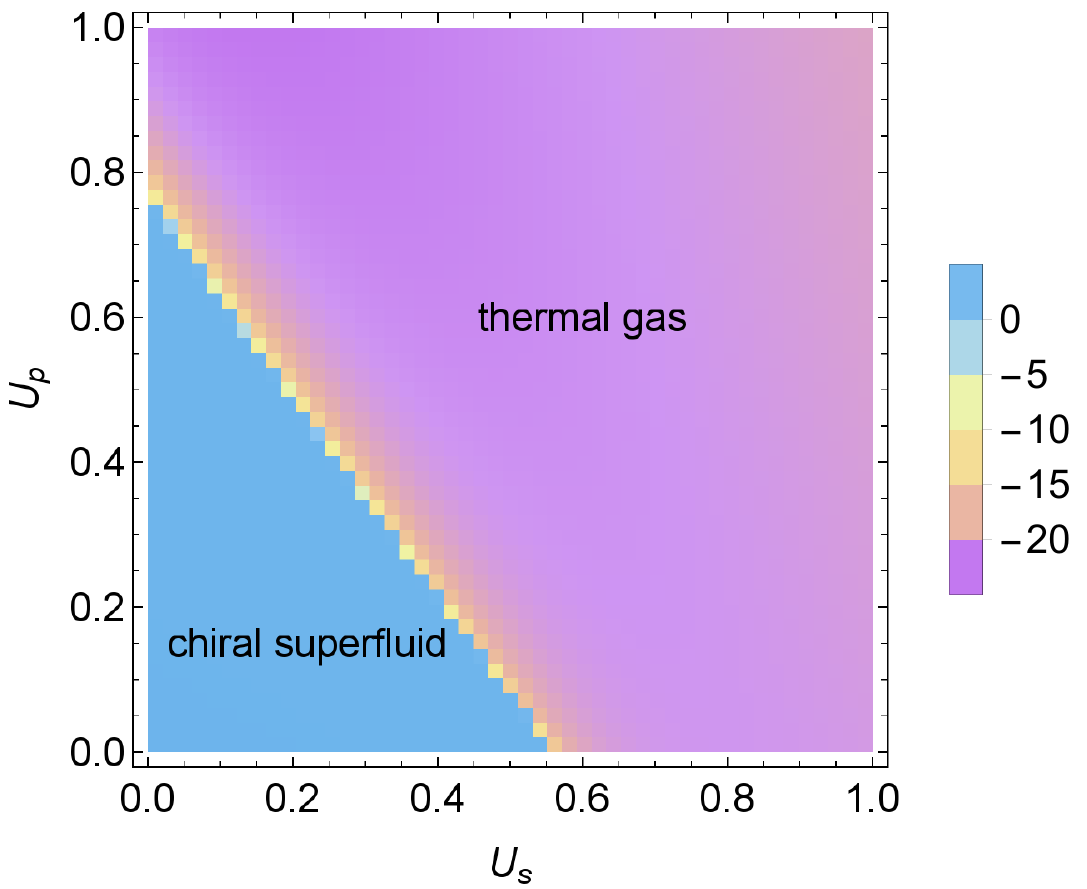}
\caption{Phase diagram for $\mu_{\Lambda}=0.2$ and $T_{\Lambda} = 0.25$. $U_s$ and $U_p$ are measured in units of $\Lambda_q^3/(2\pi^2\epsilon_\Lambda T_{\Lambda})$. The phases are determined by the sign of the chemical potential, when the RG flow is terminated at $g_u = 2$. Colors represent the final value of $\mu$. The blue region is the condensed phase separated from the thermal phase with a first-order (discontinuous) phase transition. The pink area represents the thermal phase. As $\mu_{\Lambda}$ increases or $T_{\Lambda}$ decreases, the condensed region becomes larger.}
\label{phaseD}
\end{figure}
The RG analysis can be also used to study the critical behaviors of the phase transition between the two phases. In \fref{phaseD}, we illustrate the phase diagram as a function of $U_s$ and $U_p$. The pink region represents the thermal gas phase, in which $\mu_{\Lambda}$ flow to negative values under RG transformations. Th blue region is the chiral $p_x\pm i p_y$ superfluid order with time-reversal symmetry breaking, where $\mu_{\Lambda}$ approaches one. As $\mu_{\Lambda}(0)$ increases or $T_{\Lambda}$ decreases, the phase boundary is shifted to enlarge the superfluid region. We note that along the asymptotic flow \eref{asymptotic}, the mean-field free energy can be written by two order parameters $P_{\pm} \equiv \psi_1 \pm i \psi_2$ as
\begin{equation}
F^{\Lambda}_{\rm MF} \sim \mu_{\Lambda} (|P_-|^2 + |P_+|^2) - \bar{g} (|P_-|^4 + |P_+|^4) + \mathcal{O} (P_-^6, P_+^6),
\end{equation}
where $\bar{g} >0$ is the single coefficient characterizing the asymptotic flow \eref{asymptotic}, and the last term is the higher order correction stabilizing the system. This form suggests that the transition is first order. Another indirect support for this scenario is obtained by considering the strong coupling limit for $g_0$ and $g_u$, while $g_{12}$ is set to 0. In this case, the model has effectively two XY spins on each space point that are orthogonal to each other due to the strong $g_u$ interaction. Such a model is known as Stiefel's $V_{2,2}$ model, and Monte Carlo studies show that it undergoes a first order transition \cite{Kunz1993, Loison1998, Itakura2003}. For moderate interaction strength, we speculate that the first order nature becomes weaker. Clarifying if the transition remains first order for weak interactions seems to require an extensive numerical simulations, which are beyond the scope of the paper. \footnote{The difficulty of determining the order of transitions is a common problem in the frustrated spin systems, whose effective Ginzburg-Landau models are similar to ours \cite{Kawamura1998, Itakura2003}.}

To detect this first order transition in experiments, we suggest two possibilities. The first one consists of measuring the condensate fraction as a function of temperature, preferably in a box potential \cite{Gaunt2013, Corman2014}. The measured temperature dependence will approach a non-zero jump at the condensation temperature for large systems. As a second approach we suggest to measure the spatial evolution of a phonon pulse in a condensate in a smoothly varying trap \cite{Tey2013, Sidorenkov2013, Weimer2015, Singh2016}. Here, the phonon velocity will vary as the phonon pulse approaches the interface of the condensate and the thermal gas. For a second order transition the pulse velocity will smoothly approach zero, whereas for a first order transition the velocity will approach a non-zero value before the pulse is reflected, which gives a clear indication of a first order transition.


\section{Conclusions}\label{conc}

In this paper we have investigated the critical behavior of unconventional Bose-Einstein condensates in the second band of an optical lattice. We have demonstrated that an Umklapp process between the two minima of the dispersion stabilizes a chiral superfluid state that breaks time reversal symmetry, first at the mean-field level and then within a renormalization group calculation. The latter shows that this stability is always persistent at low energy scales after integrating out thermal fluctuations. We obtain this result by identifying three separatrix planes in the RG flow, which constrain the low energy behavior to a stable regime. Furthermore, the RG flow suggests that the phase transition of the chiral superfluid state to the thermal state is of first order, in contrast to the usual second order transition of a conventional condensate. \

\

\

\ack
We thank A. Hemmerich for helpful discussions on the experimental aspects of the system. J.O., R.H., and L.M. acknowledge the support from the Deutsche Forschungsgemeinschaft (through SFB 925 and EXC 1074) and from the Landesexzellenzinitiative Hamburg, which is supported by the Joachim Herz Stiftung. W.M.H. especially acknowledges the support from Ministry of Science and Technology, Taiwan through Grant No. MOST 104-2112-M-005-006-MY3.

\appendix
\section{Effective interactions for general hopping amplitudes and interactions}\label{App}
In this appendix, we start from a single particle picture in a bipartite optical lattice to derive the hopping amplitudes and interaction parameters of a Bose-Hubbard model. We then show that the condition for the chiral superfluid state $G_1 = \tilde{g}_0 - \tilde{g}_{12} + \tilde{g}_u >0$ in \eqref{chiralC} is preserved as long as the band has two minima at $\bm{k}_1=(\pi/\sqrt{2},0)$ and $\bm{k}_2= (0,\pi/\sqrt{2})$.

\subsection{Derivation of Hubbard parameters}
\begin{figure}[!b]\center
   \includegraphics[width = 0.8\columnwidth ]{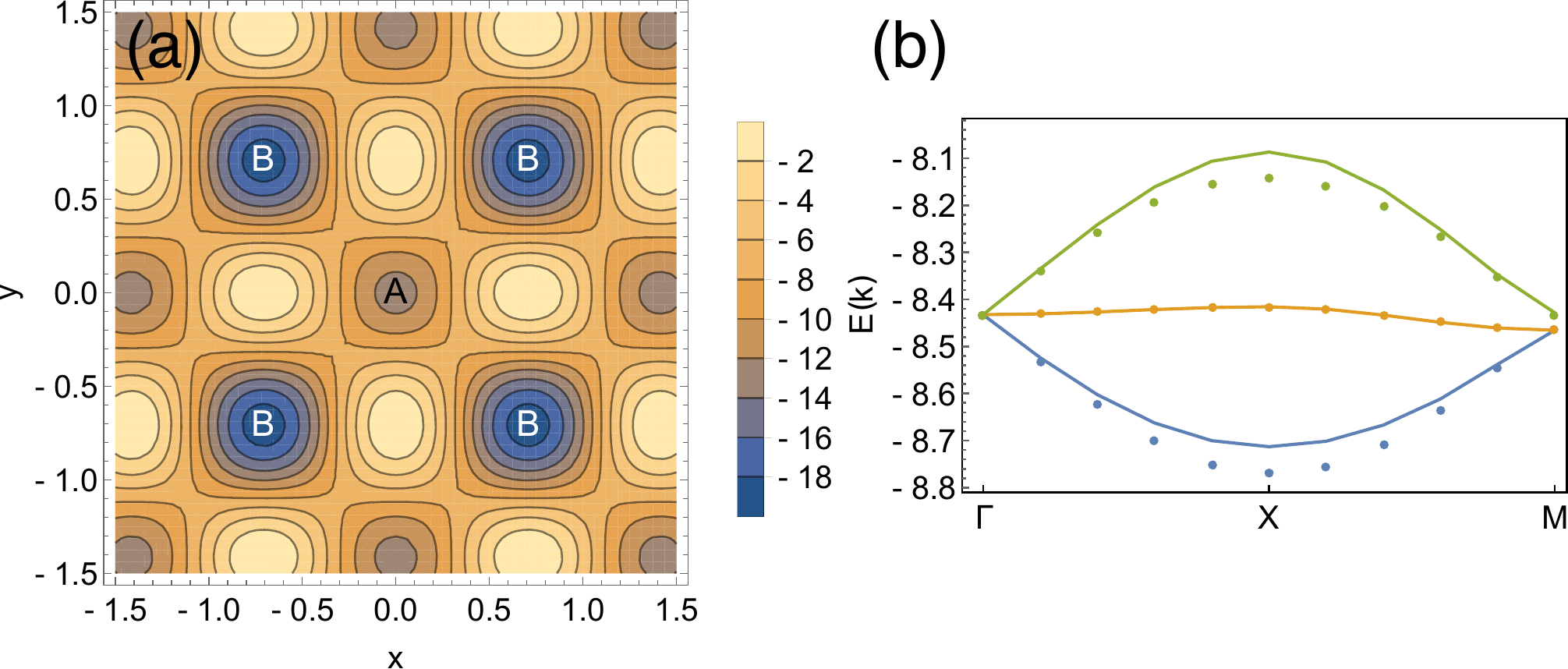}
  \caption{(a) A bipartite optical potential $V(\mathbf{r})$. (b) Numerically obtained band dispersions for the second composite bands (solid lines). Tight-binding fitting gives the dots.}
  \label{app1}
\end{figure}

We start from a following potential
\begin{eqnarray}
V(\mathbf{r}) = - V_0\left| \cos\left[  k_0 (x+y)/\sqrt{2} \right]+ e^{i\beta} \cos\left[  k_0 (x-y)/\sqrt{2} \right]  \right|^2.
\label{potential}
\end{eqnarray}
This has two local minima in a unit cell (see \fref{app1}(a)): one is a shallow local minimum hosting a $s$-orbital like Wannier state (denoted as $A$ sites), and the other is a deep minimum hosting two $p$-orbital like Wannier sates (denoted as $B$ sites). $k_0 = \pi/a$ with $a$ being the distance between neighboring $A$ and $B$ sites (\fref{lattice}(b)). The energy difference between local minima at $A$ sites and $B$ sites is $\Delta V = E_A^0 - E_B^0 = - 4 V_0\cos(\beta) $. We tune $\beta$ so that the doubly degenerate first excited states in site $B$ is close to the ground state in site $A$; in particular, we choose $\beta$ so that the three bands are exactly degenerate at the $\Gamma$ point in the following.

\begin{figure}[!tb]  \center
   \includegraphics[width = 0.6\columnwidth ]{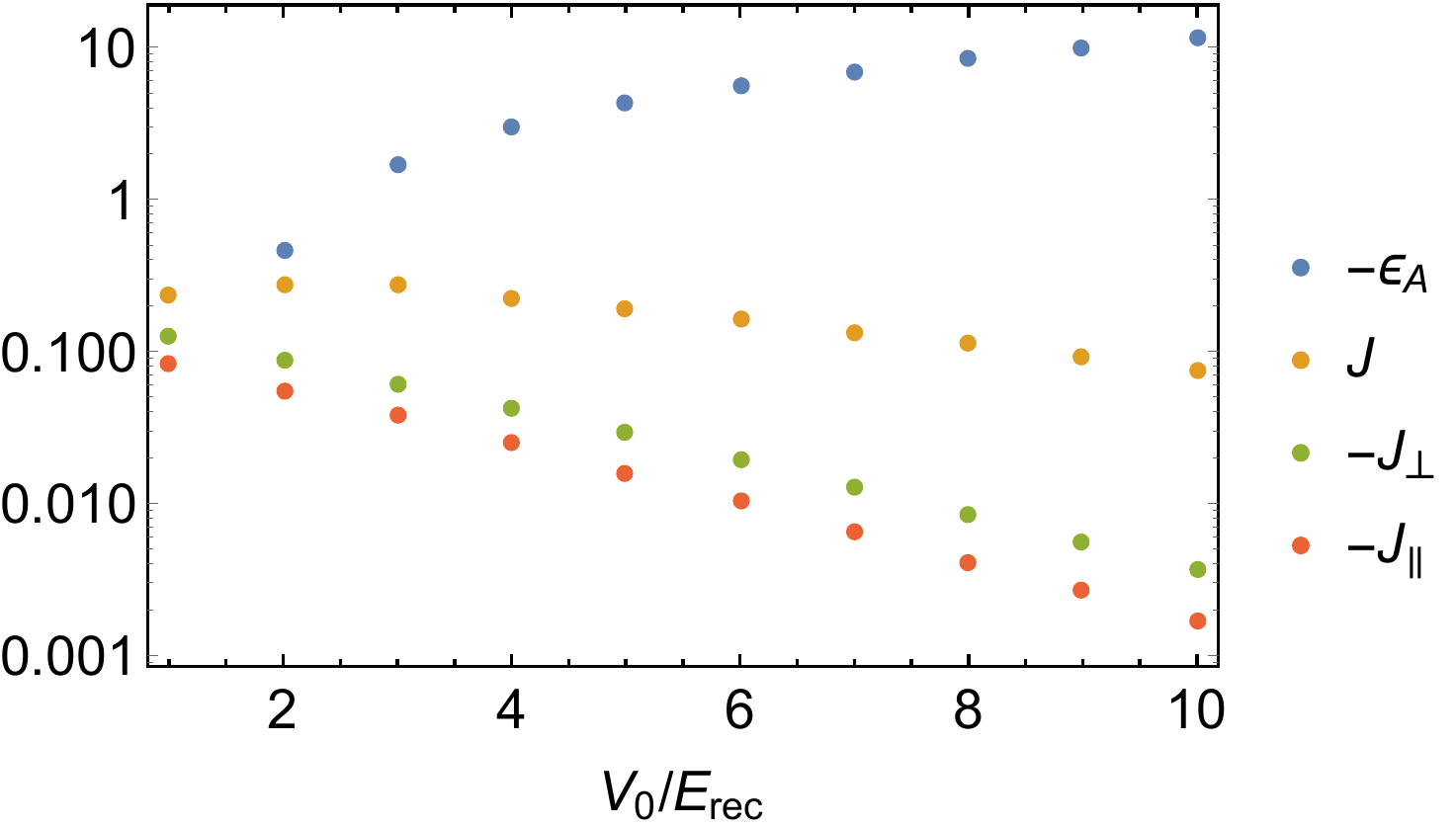}
\caption{The fitted tight-binding parameters as functions of the potential depth $V_0$ in the logarithmic scale.}
\label{app2}
\end{figure}

After solving the single particle Schr\"{o}dinger equation 
\begin{equation}
\left[ \frac{\hbar^2 \nabla^2}{2 m } + V(\mathbf{r}) \right] \psi^n_{\mathbf{k}} (\mathbf{r}) = E^n_{\mathbf{k}} \psi^n_{\mathbf{k}} (\mathbf{r}),
\end{equation}
we obtain a band dispersion as \fref{app1}(b). We fit the obtained dispersion by the following tight-binding model
\begin{eqnarray}\label{general_hopping}
\nonumber H^{\rm xy}_0 &= J \sum_{\mathbf{r} \in A} \Big[b_{1}^{\dag}(\bm{r},z)b_{2}(\bm{r}+\bm{d}_1,z)+b_{1}^{\dag}(\bm{r},z)b_{3}(\bm{r}+\bm{d}_2,z)\\
\nonumber &-b_{1}^{\dag}(\bm{r},z)b_{2}(\bm{r}-\bm{d}_1,z)-b_{1}^{\dag}(\bm{r},z)b_{3}(\bm{r}-\bm{d}_2,z)+{\rm h.c.}\Big] \\
\nonumber &- J_{\perp} \sum_{\mathbf{r} \in B, \nu = x,y} \Big[ b_{2}^{\dag}(\bm{r},z)b_{3}(\bm{r}+\bm{e}_{\nu},z)+b_{3}^{\dag}(\bm{r},z)b_{2}(\bm{r}+\bm{e}_{\nu},z)+{\rm h.c.}\Big] \\
\nonumber &- J_{\parallel} \sum_{\mathbf{r} \in B, \nu = x,y} \sum_{i=2,3}\Big[ b_{i}^{\dag}(\bm{r},z)b_{i}(\bm{r}+\bm{e}_{\nu},z)+{\rm h.c.}\Big] \\
&+ \epsilon_A \sum_{\mathbf{r} \in A} b_{1}^{\dag}(\bm{r},z) b_{1}(\bm{r},z) +  \epsilon_B \sum_{\mathbf{r} \in B} \sum_{i=2,3} b_{i}^{\dag}(\bm{r},z) b_{i}(\bm{r},z),
\end{eqnarray}
where $\bm{e}_{1} = (\sqrt{2} a,0)$ and $\bm{e}_{2} = (0, \sqrt{2}a)$. The first term is already given in \eqref{h0xy}. The degeneracy at the $\Gamma$ point is achieved by setting $\epsilon_B = \epsilon_A + 4 J_{\parallel}$. The fitted band dispersion is plotted in \fref{app1}(b) as dots. The fitted parameters as functions of the potential depth $V_0$ is given in \fref{app2}. We note that when the potential is relatively deep, the hopping between $p$-orbitals ($J_{\perp}$ and $J_{\parallel}$) is much smaller than the one between $s$- and $p$-orbitals, $J$. Thus, in practice, we can ignore $J_{\perp}$ and $J_{\parallel}$.

\begin{figure}[!tb] \center
   \includegraphics[width = 0.6\columnwidth ]{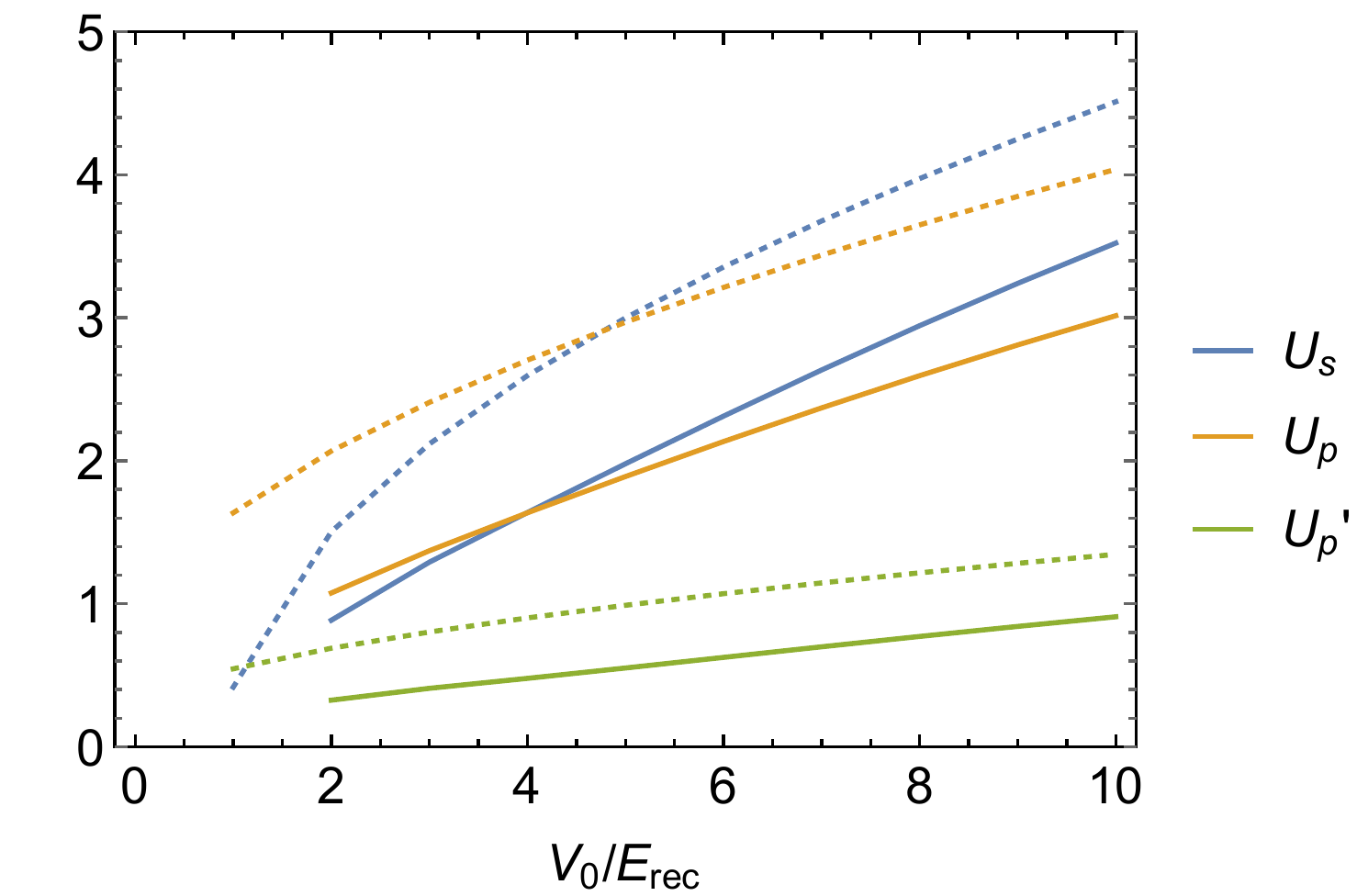}
\caption{Hubbard interactions obtained from localized Wannier functions as functions of the potential depth $V_0$. They are measured in units of $E_{\rm rec} g \int dz \left|w_z(z)\right|^4$. The dashed lines are obtained under the harmonic approximation.}
 \label{app3}
\end{figure}

Now we use the Bloch wave functions for the obtained band dispersions to construct localized Wannier functions. Here we employ a simple projection approach \cite{Marzari2012}. These Wannier functions give the bare interactions of a Bose-Hubbard model as
\begin{eqnarray}
U_s&=&g \int dz \left|w_z(z)\right|^4\int d^2{\bm r} \left|w_1(\bm r)\right|^4,\\
U_p&=&g\int dz \left|w_z(z)\right|^4\int d^2{\bm r} \left|w_{2/3}(\bm r)\right|^4, \\
U'_p&=&g\int dz \left|w_z(z)\right|^4\int d^2{\bm r} \left|w_{2}(\bm r)\right|^2\left|w_{3}(\bm r)\right|^2,
\end{eqnarray}
where $w_z(z)$ is the Wannier function of a harmonic trap along the $z$-axis, and $w_i(\bm r)$, $i=1,2,3$ are the Wannier functions of the $s$-, $p_x$ and $p_y$-orbitals on the $xy$-plane respectively. $g$ is the contact interaction strength. The obtained values are plotted in \fref{app3}. We find that $U_s \sim U_p \sim U'_p/3$ for moderately strong potential depth. 

\subsection{Effective interactions in field-theory approximations}
In this subsection, we show that the condition for the chiral superfluidity, \eref{chiralC}, is satisfied in general based on the Hubbard parameters determined above. With a general dispersion in \eqref{general_hopping} and $\epsilon_B = \epsilon_A + 4 J_{\parallel}$, the projection of the Wannier orbitals to the two minima in the lowest band becomes
\begin{eqnarray}
(u_{11}, u_{12}) &=  \frac{1}{\sqrt{2+|\lambda|^2}} \left(\lambda, \lambda^* \right),\\
(u_{21}, u_{22}) &=  \frac{1}{\sqrt{2+|\lambda|^2}} \left(1, -1\right) ,\\
(u_{31}, u_{32}) &=  \frac{1}{\sqrt{2+|\lambda|^2}} \left(1,  1\right).
\end{eqnarray}
with 
\begin{equation}
\lambda = i \frac{ \Delta_{J} - \sqrt{2 J^2 + \Delta_{J}^2}}{J} \ \rm{and} \ \it{\Delta_J = J_{\perp}- J_{\parallel}}.
\end{equation}
With \eqref{effective_int}, we can show that the 
\begin{equation}
\tilde{g}_0-\tilde{g}_{12}+\tilde{g}_u  = \frac{4  U'_p J^4  \left[2 J^2+\Delta_J^2\right] \left[ J^2- \Delta_J \left(\sqrt{2 J^2+\Delta_J^2}-\Delta_J \right)\right]}{3 \left(\Delta_J \left(\sqrt{2 J^2+\Delta_J^2}-\Delta_J \right)-2 J^2\right)^4}.
\end{equation}
For $J \gg J_{\parallel}, J_{\perp}$, the above quantity is always positive. Therefore, even when the system deviates from the simple limit of $J_{\perp} = J_{\parallel} = 0$, the condition of the chiral superfluidity is still satisfied.

\section{The $\epsilon$-expansion analysis of the fixed points}
In \sref{rg}, we have shown the fixed points that correspond to the lowest order of the $\epsilon$-expansion \cite{Janzen2016, cardy1996scaling, Wilson1972, Domany1977}. To estimate the high-order effect in the $\epsilon$-expansion, we investigate the fixed points of the RG equations without expanding $F (\mu_{\Lambda})$. Finding the zeros of the right-hand sides of \eref{muFlow}-\eref{guFlow} leads to four fixed points, except the trivial one, $(\mu^*_\Lambda,g^*_0,g^*_{12},g^*_u)=(0,0,0,0)$,
\begin{eqnarray}
\left(\mu^*_\Lambda,g^*_0,g^*_{12},g^*_u\right)&=&\left(\frac{\epsilon}{5+\epsilon},\frac{5 \epsilon}{(5+\epsilon)^2 T_{\Lambda}},0,0\right),\\
\left(\mu^*_\Lambda,g^*_0,g^*_{12},g^*_u\right)&=&\left(\frac{\epsilon}{4+\epsilon},\frac{8 \epsilon}{3(4+\epsilon)^2 T_{\Lambda}},\frac{8 \epsilon}{3(4+\epsilon)^2 T_{\Lambda}},0\right),\\
\left(\mu^*_\Lambda,g^*_0,g^*_{12},g^*_u\right)&=&\left(\frac{\epsilon}{5+\epsilon},\frac{5 \epsilon}{2(5+\epsilon)^2 T_{\Lambda}},\frac{5 \epsilon}{(5+\epsilon)^2 T_{\Lambda}},\frac{5 \epsilon}{2(5+\epsilon)^2 T_{\Lambda}}\right),\\
\left(\mu^*_\Lambda,g^*_0,g^*_{12},g^*_u\right)&=&\left(\frac{\epsilon}{5+\epsilon},\frac{5 \epsilon}{2(5+\epsilon)^2 T_{\Lambda}},\frac{5 \epsilon}{(5+\epsilon)^2 T_{\Lambda}},-\frac{5 \epsilon}{2(5+\epsilon)^2 T_{\Lambda}}\right),
\end{eqnarray}
Taking the lowest order in $\epsilon$ and setting $\epsilon = 1$ recovers the fixed points in \eref{fp1}-\eref{fp4}. We emphasize that the above fixed points still lie on the three separatrix planes $g_{u}=0$, $G_1=0$ and $G_2=0$ in the $g_0$-$g_{12}$-$g_u$ space. This fact guarantees the persistence of the chiral superfluid in the low energy limit even when we include the higher order terms in $\epsilon$ in our one-loop RG study. 

\

\

\
\providecommand{\newblock}{}

\end{document}